\documentclass[sigconf]{acmart}
\usepackage{subcaption}

\AtBeginDocument{%
  }

\setcopyright{none}
\copyrightyear{2018}
\acmYear{2018}
\acmDOI{XXXXXXX.XXXXXXX}

\acmConference[Conference acronym 'XX]{Make sure to enter the correct
  conference title from your rights confirmation emai}{June 03--05,
  2018}{Woodstock, NY}
\acmISBN{978-1-4503-XXXX-X/18/06}

\copyrightyear{2025}
\acmYear{2025}
\setcopyright{rightsretained}
\acmConference[CHI EA '25]{Extended Abstracts of the CHI Conference on Human Factors in Computing Systems}{April 26-May 1, 2025}{Yokohama, Japan}
\acmBooktitle{Extended Abstracts of the CHI Conference on Human Factors in Computing Systems (CHI EA '25), April 26-May 1, 2025, Yokohama, Japan}\acmDOI{10.1145/3706599.3719817}
\acmISBN{979-8-4007-1395-8/2025/04}

\begin{document}

\title{"Sighted People Have Their Pick Of The Litter": Unpacking The Need For Digital Mental Health (DMH) Tracking Services With And For The Blind Community}

\author{Omar Khan}
\email{mkhan259@illinois.edu}
\orcid{0009-0005-3209-3525}
\affiliation{%
    \department{Siebel School of Computing and Data Science}
    \institution{University of Illinois Urbana-Champaign}
    \city{Urbana}
    \state{Illinois}
    \country{USA}
}

\author{JooYoung Seo}
\email{jseo1005@illinois.edu}
\orcid{0000-0002-4064-6012}
\affiliation{%
    \department{School of Information Sciences}
    \institution{University of Illinois Urbana-Champaign}
    \city{Champaign}
    \state{Illinois}
    \country{USA}
}

\renewcommand{\shortauthors}{Khan et al.}

\begin{abstract}
\label{sec:abstract}

The proliferation of digital mental health (DMH) tracking services promises personalized support, yet accessibility barriers limit equal access.
 This study investigates blind community experiences with DMH tracking services across the United States as a step toward inclusive health technology design. 
Working with blind advocacy organizations, we distributed a cross-sectional observational survey (n = 93) and analyzed open-ended responses using Norman and Skinner’s eHealth Literacy framework. 
Our findings reveal significant challenges in navigation, content interpretation, and overall user experience, which impediments the blind community's effective engagement with DMH tools. Results highlight the need for adaptive interfaces, accessible tracking strategies, and voice-guided interactions. 
These insights inform design recommendations for developers and policymakers, promoting more inclusive mental health technologies. By prioritizing accessibility, we make forward progress in ensuring that DMH tracking services fulfill their potential to support mental well-being across diverse user groups, fostering digital equality in mental health care.

\end{abstract}



\received{20 February 2007}
\received[revised]{12 March 2009}
\received[accepted]{5 June 2009}

\maketitle

\section{Introduction}
\label{sec:introduction}

Over the past two decades, health and wellness technologies have rapidly evolved, shifting the reliance on strictly human evaluation of well-being toward augmenting these workflows with digital advancements.
Such advancements have been observed across multiple wellness domains, including physical health~\cite{lee_identify_2024, miller_self-monitoring_2022}, mental health~\cite{ayobi_digital_2022, oewel_approaches_2024, baghaei_time_2020}, nutritional health, sleep health, and chronic disease management~\cite{elhai_problematic_2017, kelly_its_2021}. Emerging technologies such as extended reality (XR), wearable devices, and mobile applications have shown promise in enabling more personalized and continuous monitoring of an individual's health across all these aspects of well-being~\cite{budde_arttech_2021, feinberg_zenvr_2022}. 

As defined by the World Health Organization (WHO), mental health (MH) is ```a state of well-being in which an individual realizes his or her own abilities, can cope with the normal stresses of life, can work productively and is able to make a contribution to his or her community'''~\cite{who_mh_definition}. Given the universal importance of MH, numerous technological solutions have emerged to support it, such as stress and anxiety management, mood tracking, mindfulness through meditation, medication and symptom management, and cognitive exercise. Specifically, \textit{digital mental health (DMH) tracking services} have emerged to monitor and manage mental well-being within each of these contexts. Prior work has demonstrated the benefits of these technologies, such as improved identification and understanding of thought patterns, the formation of healthier habits, and improved general well-being~\cite{bowie-dabreo_user_2022}.

However, the adoption of DMH tracking services has not been without concerns from stakeholders across the client-patient spectrum. Our study focuses on the unique concerns of the blind community, whose challenges in seeking MH support manifest themselves in both the physical and digital realms. In physical MH settings, blind individuals encounter systemic barriers, from providers’ limited cultural competency to insufficient institutional resources for effective accommodation~\cite{McDonnall02012017}. These obstacles compound in digital environments, where technical requirements such as screen reader compatibility, proper color contrast, and keyboard navigation are fundamental prerequisites for access~\cite{choi2024exploring, chaudary2023teleguidance}. The stakes are particularly high in MH contexts, where data sensitivity intersects with accessibility needs. The combination of service usability, financial constraints, and exclusion experiences creates a complex hierarchy of challenges~\cite{bowie-dabreo_user_2022}. Prior reviews have suggested a need to dive deeper into the accessibility of DMH tracking services across various populations to ensure equal access to MH support services~\cite{bunyi_accessibility_2021}. Yet, to our knowledge, there is currently limited work investigating DMH accessibility and needs with and for the blind community.

Through the design and distribution of a cross-sectional observational survey, we investigated the usage patterns of DMH tracking services throughout the blind community to develop a more sophisticated understanding of their MH needs and management strategies. The following research questions (RQs) guided our exploration:
\begin{enumerate}
    \item[\textbf{RQ1:}] What needs exist for DMH tracking services, among the blind community? 
    \item[\textbf{RQ2:}] What factors influence the adoption of DMH tracking services across the blind community?
    \item[\textbf{RQ3:}] How can data management features in DMH tracking services be improved to support data agency for blind users?
\end{enumerate}

Our survey reveals critical insights into how the blind community engages with DMH tracking services, illuminating existing and planned user preferences (RQ1), factors influencing DMH adoption, such as accessibility and unique MH goals (RQ2), and design opportunities that better facilitate blind users' MH aspirations, particularly given the sensitive nature of such data (RQ3). These findings span multiple dimensions, from the usage patterns of current DMH tracking services to specific interface challenges, data privacy concerns, and desired features for existing and future platforms. We provide a foundation for understanding how DMH tracking services can better serve blind users in alleviating obstacles to MH care through enhanced accessibility considerations, and more importantly, designing \textit{with} and not exclusively \textit{for} this community.

We first situate our work within the broader landscape of DMH services, examining their evolution and current limitations in supporting diverse user needs (Section~\ref{sec:related_work}). We then detail our approach to understanding blind users’ experiences with DMH tracking services, including our participant recruitment, survey design, and analysis procedures (Section~\ref{sec:methods}). Our findings (Section~\ref{sec:findings-discussion}) reveal community-driven insights about barriers to receiving MH care exacerbated by DMH tracking services, leading to a set of design implications for creating more inclusive MH technologies. We conclude by outlining a research agenda for advancing accessible DMH tracking services for the blind community and providing concrete recommendations for researchers and practitioners (Section~\ref{sec:conclusion}).

\section{Related Work}
\label{sec:related_work}

This section outlines previous contributions to the advancement of MH in HCI and what remains to be explored for access to the inclusive DMH tracking service. 

\subsection{Mental Health Research in Human-Computer Interaction}
\label{subsec:mh_in_hci}

HCI has increasingly focused on health and wellness technologies, especially mental well-being~\cite{coyle_interaction_2012, calvo_design_2019}. Prior work has explored workplace stress reduction~\cite{chapman_2019_it}, academic anxiety~\cite{kelly_its_2021}, social connectivity's effect on MH~\cite{balcombe_evaluation_2023}, and issues faced by marginalized groups~\cite{ayobi_digital_2022}. Emerging technologies like LLM chatbots for cognitive therapy~\cite{gu_mentalblend_2024, li_automatic_2024}, VR exposure therapy~\cite{baghaei_time_2020, feinberg_zenvr_2022, meinlschmidt_mental_2023}, and smartphone mood tracking~\cite{alslaity_insights_2022, hoefer_visualizing_2022} have also been explored.

These innovations offer benefits like improved availability and engagement in MH care~\cite{bowie-dabreo_user_2022, bunyi_accessibility_2021} and personalized MH goal-setting ~\cite{zhang_designing_2021}, but also presents concerns ~\cite{hassan_unveiling_2023, kang_this_2024, khoo_thats_2024, oguamanam_intersectional_2023, robledo_yamamoto_therapy_2021}. In HCI, limited exploration of their effectiveness and accessibility has been done for diverse populations, especially those with disabilities ~\cite{bunyi_accessibility_2021, sien_designing_2023}. Our work begins to address these gaps, focusing on disabled users.

The rise of DMH tracking services significantly enhances MH care due to a wide range of user support. These services include mood monitoring~\cite{barry_mhealth_2017}, guided meditation and mindfulness~\cite{dauden_roquet_evaluating_2018, markum_digital_2020}, and peer support communities~\cite{oleary_suddenly_2018}. They have evolved from simple self-help apps to advanced AI-driven platforms offering personalized interventions~\cite{ausman_artificial_2019, cabrera_ethical_2023}. DMH services reduce barriers to access support~\cite{adams_availability_2024, lattie_overview_2022}, offer timely help~\cite{bowie-dabreo_user_2022, calvo_positive_2017}, and enable continuous MH monitoring~\cite{kornfield_energy_2020, nepal_current_2021}. However, their effectiveness compared to traditional in-person interventions is still debated~\cite{cabrera_ethical_2023, robledo_yamamoto_therapy_2021}. Concerns about user engagement, adherence, and potential negative effects have been reported, including financial inaccessibility among prenatal Black women of lower socioeconomic status~\cite{oguamanam_intersectional_2023} and reported impracticalities associated with carrying extra devices on d daily basis ~\cite{sheikh_wearable_2021}.

\subsection{DMH Tracking Services And The Blind Community}
\label{subsec:dmh_and_blind}

The intersection of accessibility and MH technology is emerging yet understudied in computing and related fields. While DMH services can democratize access to MH support, their accessibility for disabled users, especially the blind community, is not well understood~\cite{bunyi_accessibility_2021, kohda_mental_2019}. This is concerning as blind individuals often face unique MH challenges and higher rates of depression and anxiety compared to the general population~\cite{kohda_mental_2019, richardson_underutilization_2024}.

Recent research has explored the needs and preferences of blind individuals in personal health tracking~\cite{lee_identify_2024, miller_self-monitoring_2022, choi2024exploring}. Studies emphasize adaptive interfaces~\cite{hoefer_visualizing_2022, slovak_hci_2024}, audio feedback~\cite{chapman_sociotechnical_2024, rector_exploring_2015}, and haptic interactions to convey emotional information~\cite{rector_exploring_2015, soler-dominguez_arcadia_2024}. However, studies on the effectiveness of current DMH platforms for the blind are limited. Understanding the usage of DMH tracking tools for blind individuals helps identify accessibility gaps in DMH services~\cite{calvo_computing_2016, calvo_design_2019, lattie_overview_2022}, leverage technology for customized interventions~\cite{miller_self-monitoring_2022, pandey_mental_2023}, and contribute to broader accessibility research~\cite{sien_co-designing_2023, sien_designing_2023, slovak_hci_2024}. Despite progress in HCI for MH and personal health (PH) tracking, there's a need to examine these technologies' use and accessibility for the blind. Our research aims to address this by understanding and improving DMH services' accessibility and efficacy for blind individuals, offering a direction for inclusive DMH solutions.
\section{Methods}
\label{sec:methods}

\begin{figure*}[ht!]
    \centering
    \includegraphics[width=\textwidth]{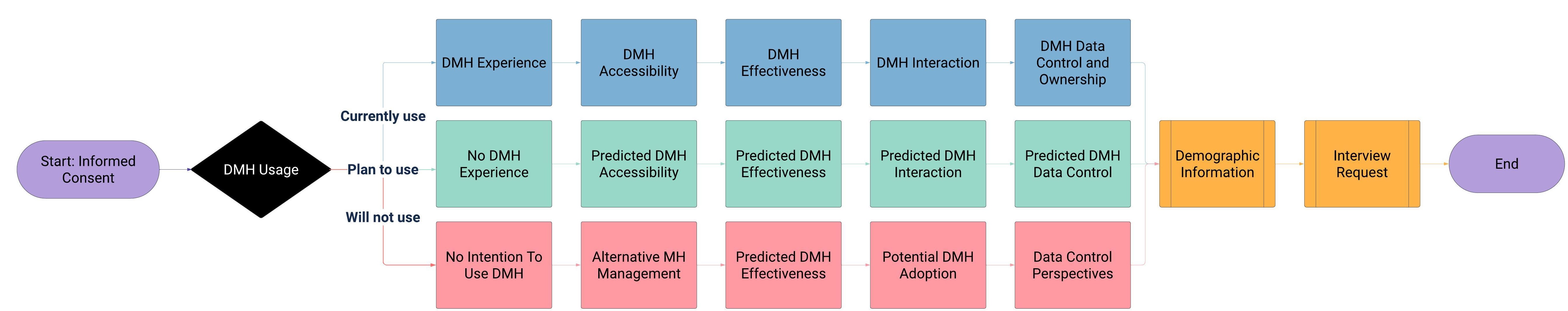}
    \Description[DMH Survey Flow Disgram]{Flow chart depicting the three tracks that survey respondents could have taken. The diagram starts with 'Informed Consent' and branches into three paths based on DMH Usage: 'Currently use', 'Plan to use', and 'Will not use'. Each path progresses through stages of Experience/No Experience, Accessibility, Effectiveness, Interaction, and Data Control. The 'Currently use' and 'Plan to use' paths converge at 'DMH Data Control and Ownership' and 'Predicted DMH Data Control' respectively. Before ending, all paths lead to 'Demographic Information' and 'Interview Request'. The chart uses color coding: blue for current users, green for planned users, and maroon for non-users.}
    \caption{Flow diagram displaying the survey's flow. Participants received varying questions depending on their response to the survey's "DMH Usage" portion. Upon completing the DMH-related questions, all participants had the chance to complete the same demographic and interview request questions.}
    \label{fig:survey-flow}
\end{figure*}

\subsection{Research Team}
\label{subsec:research-team}


To establish methodological transparency, we begin by acknowledging our positionality as researchers and its influence on our study design and analytical framework. The research team comprises two blind authors: the first author is low-vision, while the second author is totally blind. This experiential diversity within the research team enhanced our methodological approach by incorporating multiple perspectives on blindness, thereby strengthening our capacity to comprehensively examine how varying degrees of blindness impact users' engagement with DMH tracking services. Our distinct lived experiences informed both the conceptual framework and interpretative lens of this study, aligning with established practices in disability studies research that value disabled researchers' positionality as a methodological asset~\cite{mankoffDisabilityStudiesSource2010, sharifShouldSayDisabled2022}.

\subsection{Study Design} 
\label{subsec:study-design}


Several factors played a crucial role in choosing the methodological approach for this study. The primary challenge was ensuring sufficient breadth and depth in responses to accurately capture the current use of MH technology by the blind community. Moreover, achieving data saturation with this population can be challenging, as such a pool is not readily accessible~\cite{Smith2018}. To capture a breadth of insights into whether the blind community engages with DMH tracking services and to what extent, we designed a survey that took inspiration from several existing questionnaires, including the Technology Acceptance Model (TAM)~\cite{davis1989technology} and the Client Satisfaction Questionnaire (CSQ)~\cite{attkisson_client_1982}. To curate the list of DMH tracking services we would be asking about, we took inspiration from several existing frameworks and taxonomies for classifying DMH interventions~\cite{pineda_updated_2023, chen_hybrid_2024, cross_digital_2024}. This study design was advantageous for our RQs, as we aimed to capture both breadth and \textit{some} amount of depth in DMH service usage, and a survey serves as a strong starting point to meet these goals. It also allowed us to investigate a historically well-defined problem: the need for diverse voices when designing DMH services~\cite{kohda_mental_2019, bunyi_accessibility_2021}.
We designed a cross-sectional observational survey on Qualtrics~\cite{qualtrics} to gather insights on blind individuals' current use of DMH tracking services, including close-ended questions asking about typical usage patterns of these services and types of services used (RQ1), a mixture of open-ended and close-ended questions asking about factors influencing DMH adoption (RQ2), and open-ended questions asking participants to reflect on data management practices with DMH tracking services (RQ3). Figure~\ref{fig:survey-flow} provides a visual of the different completion pathways of our survey. 

\subsection{Participant Recruitment}
\label{subsec:recruitment}

To build a comprehensive sense of the challenges faced by the blind community with DMH tracking services, we ensured that our study captured a holistic view of accessibility challenges and opportunities within this domain. Thus, our inclusion criteria included the following: \textbf{(1)} The participant must be 18 years of age or older, \textbf{(2)} The participant must identify as legally blind, and \textbf{(
3)} The participant must be physically located in the United States. The participant pool was limited to the United States for this initial investigation to eliminate international payment processing complications and streamline survey timing and deployment.

We collaborated with several blind advocacy organizations within the United States, including the National Federation of the Blind (NFB)~\footnote{https://nfb.org/}, the American Council of the Blind (ACB)~\footnote{https://www.acb.org/}, the American Foundation for the Blind (ACB)~\footnote{https://www.afb.org/}, and the DO-IT mailing lists managed by the University of Washington~\footnote{https://www.washington.edu/doit/}. Our university's Institutional Review Board (IRB) reviewed this study and marked it exempt. Following the review and approval of our community partners, they distributed our study materials via their mailing lists. Survey deployment began in August 2024, and 124 responses have been collected to date.

\subsection{Survey Analysis}
\label{subsec:survey-analysis}

We implemented a comprehensive analytical framework that combined statistical methods for closed-ended questions with systematic qualitative analysis of open-ended questions (survey materials are provided in Appendix, Section~\ref{sec:survey-instrument}). Our closed-ended response analysis used a frequentist lens, chosen for its ability to provide objective results without requiring prior assumptions~\cite{fornacon-wood_understanding_2022} - particularly crucial given the limited existing research on blind individuals’ experiences with DMH tracking services. We conducted descriptive statistics for all closed-ended items, including measures of central tendency and variability. To ensure statistical rigor, we employed the Kruskal-Wallis test~\cite{kruskal1952use} to examine differences in the perceived effectiveness of different service categories. We followed a systematic coding protocol for open-ended responses using ATLAS.ti 24 for Mac~\cite{ATLASTI_2024}. This process involved initial open coding, axial coding to identify relationships between concepts, and selective coding to map open-ended responses to the appropriate facet(s) within the eHealth literacy framework~\cite{norgaard2015health}. 

\section{Findings and Discussion}
\label{sec:findings-discussion}
\begin{figure*}[ht]
    \centering
        \includegraphics[width=\textwidth]{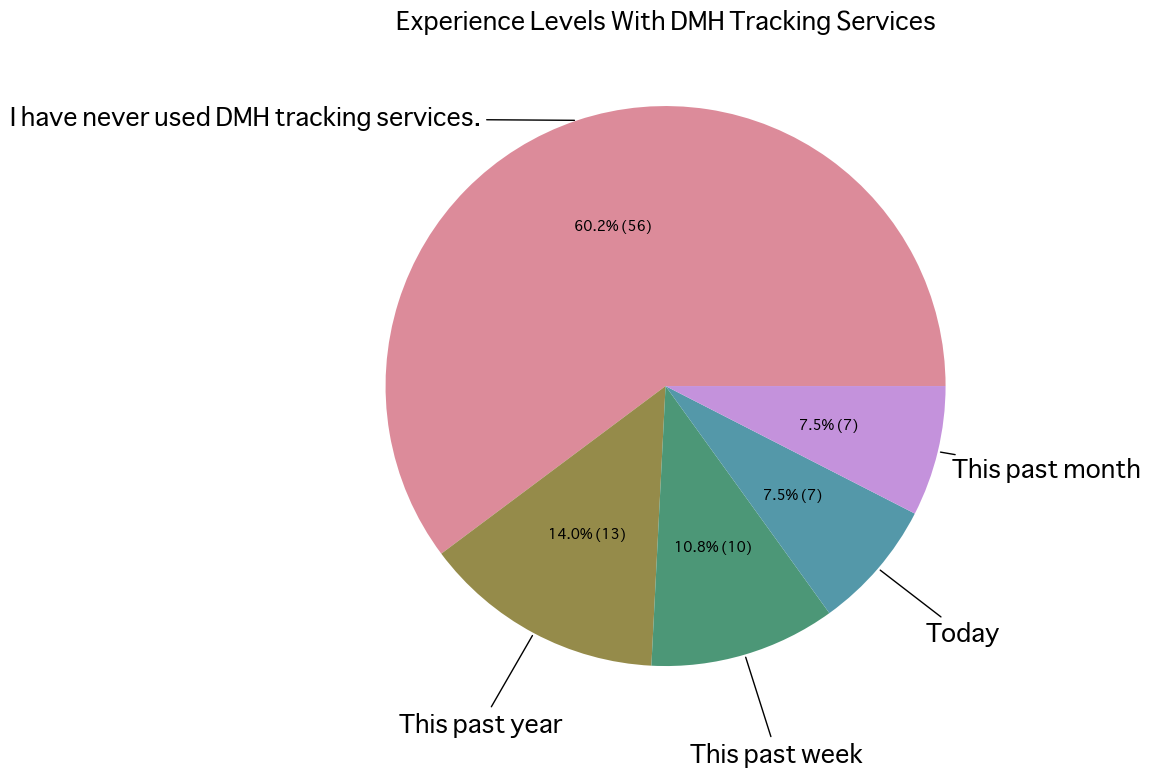}
        \caption{Participants' reported experience levels with DMH tracking services (n=93).}
        \Description{Pie chart titled ‘Experience Levels With DMH Tracking Services’. The chart shows the following segments: ‘I have never used DMH tracking services’ at 60.2\% (56 responses), ‘This past year’ at 14.0\% (13 responses), ‘This past week’ at 10.8\% (10 responses), ‘Today’ at 7.5\% (7 responses), and ‘This past month’ at 7.5\% (7 responses).}
    \label{fig:exp-levels}   
\end{figure*}


Our analysis of participants' survey responses revealed critical insights about usage patterns, accessibility barriers, and desired features in current DMH tracking services. We mapped open-ended responses to Norman and Skinner's eHealth literacy framework~\cite{norgaard2015health}, identifying significant gaps in several types of literacy. Our analysis addresses three key areas: needs assessment (RQ1), adoption factors (RQ2), and data management improvements (RQ3) for accessible DMH tracking services.

\subsection{The Need For DMH Services Across The Blind Community}
\label{subsec:dmh-service-needs}


\begin{figure*}[ht]
    \centering
        \includegraphics[width=\textwidth]{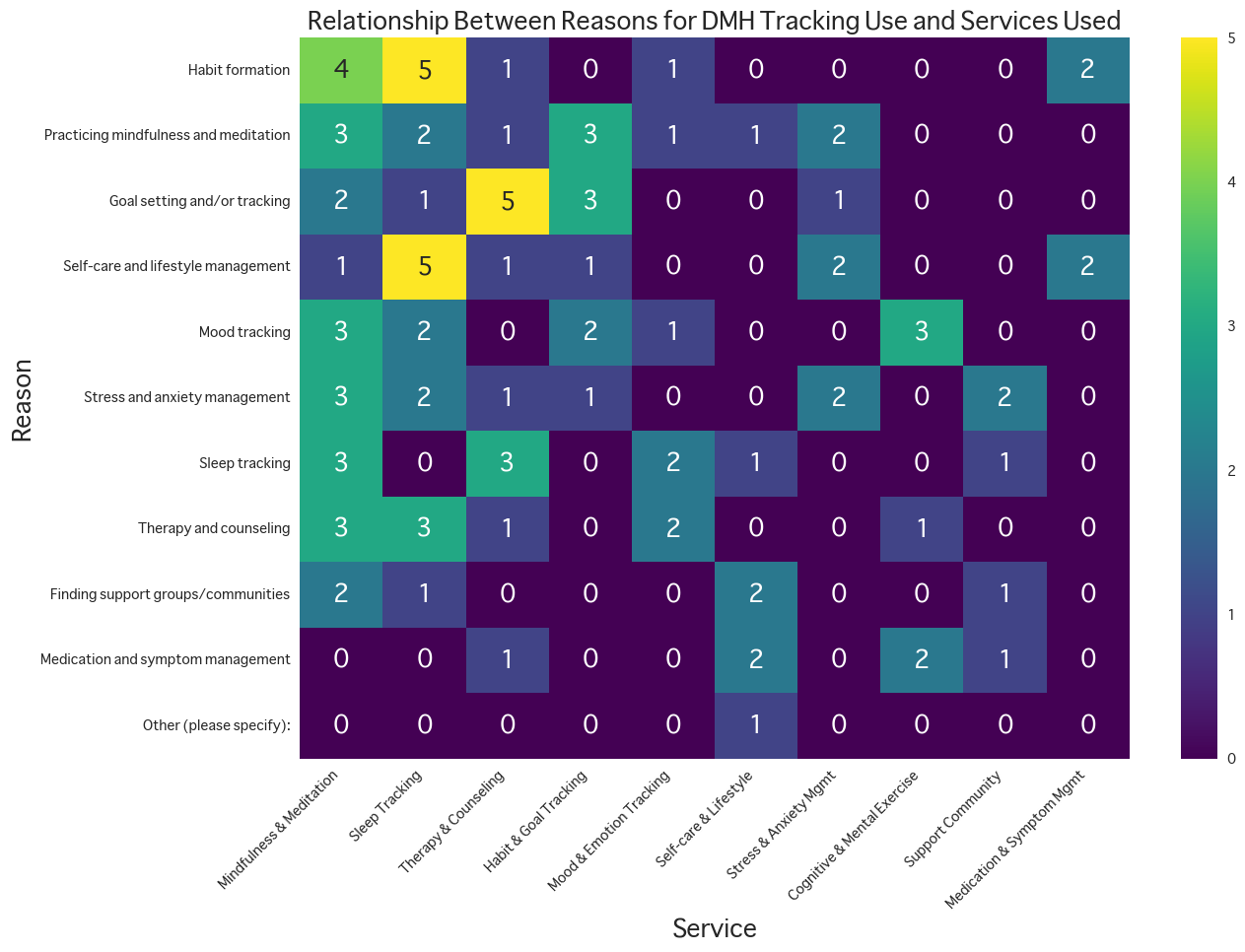}
        \caption{Heatmap illustrating the relationship between DMH tracking service usage frequency and reasons for use across the blind community.}
        \Description{A heatmap visualization showing the relationship between reasons for digital mental health (DMH) tracking use (y-axis) and services used (x-axis). The heatmap uses a color scale from dark purple (0) to bright yellow (5) to represent frequency or strength of relationships. The y-axis lists 11 reasons including habit formation, mindfulness practice, goal setting, self-care, mood tracking, stress management, sleep tracking, therapy support, finding support groups, medication management, and other. The x-axis shows 10 service categories including mindfulness \& meditation, sleep tracking, therapy \& counseling, and various management tools. The visualization reveals varying intensities of relationships between different reasons and services, with some notable clusters of higher values (yellows and greens) in certain intersections.}
    \label{fig:reasons-services-relationship}
\end{figure*}

In total, 124 participants attempted the survey and 93 participants (75\%) fully completed the survey. Of those who completed the survey, the median completion time was 17.64 minutes ($s$ = 16.2 minutes; $Q_{1}$ = 12.71 minutes; $Q_{3}$ = 24.25 minutes). Table~\ref{tab:survey-demographics} includes demographic characteristics of the final dataset used for analysis. 
Addressing RQ1, our survey findings revealed diverse DMH tracking service needs within the blind community. However, mindfulness/meditation services and sleep tracking emerged as the most popular among blind users. Figure~\ref{fig:exp-levels} provides an overview of experience levels with DMH tracking services, indicating that indicates that 60.2\% of participants have never used DMH tracking services, while a smaller proportion have used them in the past year (14\%), past week (7.5\%), past month (7.5\%), or today (10.8\%). Figure~\ref{fig:reasons-services-relationship} shows the distribution of currently used service categories, showing that “Habit formation” and “Self-care \& lifestyle management” are the most commonly reported reasons for using services like Mindfulness \& Meditation and Therapy \& Counseling, while other reasons like “Medication \& symptom management” are less frequently associated with these services.

Participants gave varied reasons when asked to elaborate on these services' need. For example, P29 described their reliance of several DMH tracking services: "\textit{They helped me keep track of my medications, various doctor's appointments. I also have an app that I [use] to help me with my anxiety, helps me track my sleep cycle through my Apple Watch as well...send information if my doctor would like it...can be forwarded to them...and I’ve been able to reduce the psychiatric medication that I was once on}" - \textit{P29}. Additionally, P70 dives deeper into their use of meditation apps and how they provide them with guided strategies for relaxation: "\textit{...I find it to be very relaxing, especially with the guided meditations, and the great music choices they have available! This is true before bed, in the middle of the afternoon, etc.}" - \textit{P70}. These initial accounts suggest that the blind community may favor mindfulness and sleep tracking services for their positive lifestyle impacts, suggesting important implications for future DMH tracking services.

While existing literature has not systematically mapped blind users’ preferences for specific MH app categories, our participants’ accounts reveal popular MH management strategies for this community. These user-driven insights should guide future DMH tracking services, shifting from retrospective accessibility adaptations to proactively inclusive design approaches.

\subsection{Factors Influencing DMH Adoption Across the Blind Community}
\label{subsec:survey-factors}


\begin{figure*}[t!]
    \centering
    \begin{subfigure}[t]{0.3\textwidth}
        \centering
        \includegraphics[width=\textwidth]{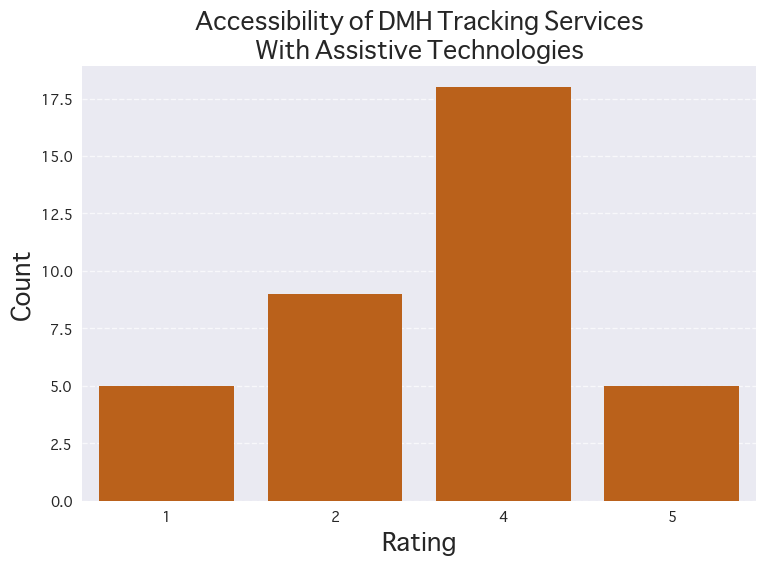}
        \caption{Distribution of participant ratings for accessibility. Ratings were measured on a 5-point Likert scale (1 = \textit{Very inaccessible}, 3 = \textit{Neither accessible nor inaccessible}, 5 = \textit{Very accessible}).}
        \Description{Bar chart displaying accessibility ratings of DMH Tracking Services with assistive technologies. X-axis shows ratings from 1-5, Y-axis shows count of responses. Distribution: Rating 1: 5 responses, Rating 2: 9 responses, Rating 4: 18 responses, Rating 5: 5 responses. Note: Rating 3 has no data.}
        \label{subfig:access}
    \end{subfigure}
    \hfill
    \begin{subfigure}[t]{0.3\textwidth}
        \centering
        \includegraphics[width=\textwidth]{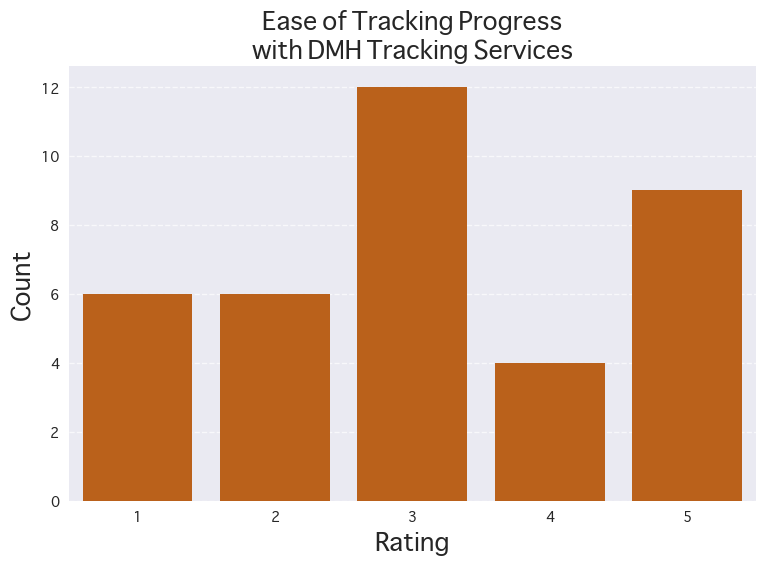}
        \caption{Distribution of participant responses to tracking progress difficulty. Ratings were measured on a 5-point Likert scale (1 = \textit{Very difficult}, 3 = \textit{Neither easy nor difficult}, 5 = \textit{Very difficult}).}
        \Description{Bar chart showing distribution of ratings for ease of tracking progress with DMH tracking services. X-axis shows ratings from 1-5, Y-axis shows count of responses. Distribution: Rating 1: 6 responses, Rating 2: 6 responses, Rating 3: 12 responses, Rating 4: 4 responses, Rating 5: 9 responses.}
        \label{subfig:progress}
    \end{subfigure}
    \hfill
    \begin{subfigure}[t]{0.3\textwidth}
        \centering
        \includegraphics[width=\textwidth]{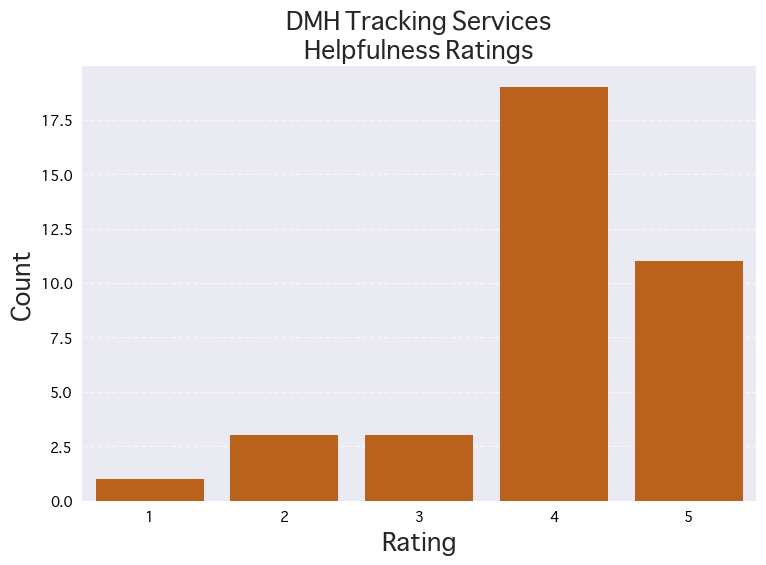}
        \caption{Distribution of participant ratings towards DMH helpfulness. Ratings were measured on a 5-point Likert scale (1 = \textit{Very unhelpful}, 3 = \textit{Neither helpful nor unhelpful}, 5 = \textit{Very helpful}).}
        \Description{Bar chart illustrating helpfulness ratings for DMH tracking services. X-axis shows ratings from 1-5, Y-axis shows count of responses. Distribution: Rating 1: 1 response, Rating 2: 3 responses, Rating 3: 3 responses, Rating 4: 19 responses, Rating 5: 11 responses.}
        \label{subfig:helpfulness}
    \end{subfigure}
    
    \caption{Participant ratings for DMH tracking services helpfulness and accessibility.}
    \Description{Three bar charts showing participant ratings for DMH tracking services' helpfulness, ease of tracking progress, and accessibility via assistive technologies arranged in a row.}
    \label{fig:dmh-distributions}
\end{figure*}

Turning to RQ2, our findings reveal how MH tracking motivations among blind users intersect with accessibility demands, creating tension between tracking priorities and tool adoption. While sharing core tracking needs with non-blind users, blind individuals encounter distinct barriers to usage, resulting in a disproportionate user burden.
Figure~\ref{fig:reasons-services-relationship} depicts the most common MH-related motivations behind each DMH category adoption. When asked to elaborate on these responses, some stressed the importance of \textit{mentorship and awareness for engagement}. P13, for instance, cited their occupation as a MH counselor and the need to be familiar with these services: "\textit{I have a masters in clinical mental health counseling and have explored them to recommend to clients}" - \textit{P13}. This shows interest in both \textit{personal} and \textit{professional} uses of DMH tracking services within the blind community, with some exploring these services beyond basic use. Others highlighted more MH-specific reasons, such as P61 who wanted to "\textit{Improve sleep quality and duration}" - \textit{P61}. Participants also valued the ability to self-pace MH goals as noted by P32 and how: "\textit{[they] allow you to proceed at your own pace}" - \textit{P32}, as well as P52: "\textit{accountability for myself and that others could see my progress to encourage me}." - \textit{P52}.   

Design factors played an equally key role in DMH adoption, particularly \textit{accessibility} and \textit{usability} of these services. Figure~\ref{subfig:access} shows participants' closed-ended responses when asked "\textit{How accessible are digital mental health tracking services for you when using tools like assistive technologies?}" and Figure~\ref{subfig:access-issues} shows their various reasons for these ratings. When asked to elaborate on any additional accessibility challenges, participants expressed several accessibility shortcomings. P8 described current DMH tracking services as: 
\begin{quote}
    "\textit{Mostly unhelpful because \textbf{they are at times not accessible with a screen reader.} Furthermore, they require me to go out of my way to...set them up, and...learn their UI/UX which at times is significantly different from what I am used to with my device}" - \textit{P8}. 
\end{quote}

Another obstacle was the lack of assistive technology (AT) support, as expressed by P25: "\textit{A lot of these apps don’t even allow for the use of adaptive screen reader technology, and various apps that assist people with vision loss}" - \textit{P25}. Participants also noted convoluted user experience, specifically finding the desired feature with ease. P67 voiced their concerns with one app that they used: "\textit{...I'd encounter [unlabeled] buttons, screens and interfaces that weren't navigable on Voiceover, and so much more.  I've even encountered apps where couldn't even get past the startup/welcome screens! I've also seen it where creating accounts was accessible, but, the rest of it was not...I'd need to use so many workarounds, that it would defeat the purpose of the app}" - \textit{P67}. Prospective users (n=29) echoed these concerns, as seen in Figure~\ref{subfig:prereq-features} which highlights several prerequisites to DMH adoption. Feedback also emphasizes the need to reduce cognitive load when learning new technologies and MH strategies; such increased cognitive was noted because of platform inaccessibility, as suggested by P65: 

\begin{quote}
    "\textit{What made them ineffective, was the lack of accessibility, making them either unuseable, or very hard to use with Voiceover which, I need, as it's my screenreader of choice. The list of issues is endless, from unlabeled buttons, to elements that weren't seen by voiceover, etc...\textbf{I don't have a lot of is any choices of what apps or services I use, while sighted people, have their pick of the litter}}" - \textit{P65}.
\end{quote}

Participants also indicated the importance of \textit{accessible progress tracking and data visualization}; Figure~\ref{subfig:progress} shows distributions of responses when asked "\textit{How difficult is it for you to track your progress towards achieving mental health goals using digital mental health tracking platforms?}" and Figure~\ref{subfig:obstacles} highlights aspects of current progress tracking implementations that pose the most obstacles to current users. Interpreting visual data is a major concern, consistent with previous studies on data visualization accessibility for the blind community. Furthermore, \textit{financial} factors also heavily influenced DMH adoption, as expressed by P75 surrounding their own DMH selection and adoption, noting that: "\textit{Most have a huge cost associated with them that doesn't seem worth it, and many have accessibility challenges or glitches sprinkled throughout them}" - \textit{P75}.

These adoption criteria span MH readiness, design, and financial barriers, aligning with recent DMH research and underscoring the imperative for diverse stakeholder involvement in health technology design~\cite{waqqas_khokhar_burgeoning_2024, jovanovic_barriers_2024, calvo_computing_2016, calvo_design_2019}. Through the lens of the eHealth Literacy framework~\cite{norgaard2015health}, our findings highlight critical computer and information literacy gaps as blind users navigate complex interfaces. This suggests the need to shift from retrofitted accessibility to proactively inclusive design approaches that address eHealth literacy dimensions from inception. This may include robust screen reader support and clutter-free UI design (e.g., reducing visual noise, creating a clear content hierarchy, leveraging progressive disclosure~\cite{smith2023}). 

\subsection{The Blind Community's Data Agency Preferences with DMH Tracking Services}
\label{subsubsec:suvey-dmh-data-agency}


Investigating RQ3, participants expressed various concerns about the security and privacy of their data with DMH tracking services, highlighting the blind community's unique vulnerability to insecure systems. 
One recurring concern was \textit{confidentiality}. In particular, P10 expressed their concern about untrained MH providers having access to their data: "\textit{I do not like the idea of people who are not related to my medical care having access to sensitive information}" - \textit{P10}. P53 likened the importance of such security standards to traditional health contexts: "\textit{It needs to be safe and there must be confidentiality practice just like [in-person] mental health counseling}" - \textit{P53}. Confidentiality was also a significant factor for those who indicated no desire to use DMH tracking services in the future, as said by P7:
"\textit{[Confidentiality] important to me because there is still a stigma around mental health, chronic health, disability, etc. And I do not want my medical information shared with others without my consent because that information could be viewed negatively and used against me}" - \textit{P7}. P7's account perpetuates documented patterns of MH stigmatization, extending prior work on discriminatory data practices.

Another concern was the \textit{lack of agency} for data management and sharing, as highlighted by P39: "\textit{I am a human and have the right to control/manage my own affairs just like every other individual on this earth whether or not they have a disability.}" - \textit{P39}. P39's account is particularly noteworthy due to its emphasis on disability, and that having agency in one's life is a universal right. Moreover, participants reiterated the importance of \textit{transparency} when using DMH tracking services and how all data management measures should be clear from the start of service usage. P12 emphasized this as they detailed their thoughts on data management: "\textit{I want to know who is seeing my information but at the end of the day I need the tools that I need so will share some things. Not really hiding anything}" - \textit{P12}. 

These privacy concerns echo established findings on MH data sensitivity and practitioner confidentiality~\cite{Anvari2024, Progga2025}. They also reinforce prior work on security and privacy challenges unique to the blind community when using health technologies~\cite{lee2023personal}. Under Norman and Skinner’s eHealth literacy framework~\cite{norgaard2015health}, our participants’ responses reveal particular challenges with information literacy and media literacy - specifically, understanding how their MH data is stored, shared, and protected in digital environments. This suggests that DMH services must not only implement robust privacy measures but also communicate these protections in accessible formats, helping users develop the literacy skills needed to make informed decisions about their MH data management.

\subsection{Limitations}
\label{subsec:limitations}

While our survey captured broad DMH tracking trends across platforms, this approach limited platform-specific insights and lacked integration of validated tools like TUQ~\cite{parmanto_development_2016} and CSQ~\cite{attkisson_client_1982}  for measuring telehealth usability and satisfaction. The dynamic DMH landscape suggests value in focused studies of specific services or demographics, particularly through longitudinal analysis. Moreover, as this was an observational survey and not an experimental survey, we cannot guarantee any generalizability, and further studies must be conducted to scale up to the broader blind and low-vision (BLV) community.Our cross-sectional methodology constrained temporal understanding of user engagement patterns and adaptation over time. Future work will employ Creswell’s explanatory sequential mixed-methods design~\cite{creswell2017designing} to complement quantitative findings with qualitative depth, enabling richer insights into user experiences and service adoption barriers.

\section{Conclusion}
\label{sec:conclusion}

Our survey of 93 blind participants revealed both high demand for specific DMH features (particularly mindfulness promotion and sleep tracking), significant accessibility barriers, and data management concerns. Analysis through Norman and Skinner’s eHealth literacy framework~\cite{norgaard2015health} highlighted critical gaps in several literacies that impede the effective use of current DMH services. These findings suggest an urgent need for DMH tracking services that both address blind users’ documented preferences and support the development of essential eHealth literacies. Future work will explore these insights through stakeholder interviews to inform more inclusive DMH solutions.
\begin{acks}
We thank all community partners for their assistance in distributing our study materials to their respective members and for their enthusiasm and support of this work. We would also like to thank our research group for their feedback throughout the entire study. 
\end{acks}

\bibliographystyle{ACM-Reference-Format}
\bibliography{refs,a11y_framework}

\section{Appendix}
\label{sec:appendix}
\appendix

\section{Preliminary Findings}
\label{sec:prelim-findings}

\newlength{\tab}
\setlength{\tab}{0.1em}

\begin{table}[htbp]
\centering
\begin{tabular}{ll}
    \toprule
    \textbf{Demographic} & \textbf{n}\\
    \hline
    \textbf{Participants} & \textbf{93}\\
    \hline
    \textbf{Gender} & \\
    \hspace{\tab} Woman & 64 \\
    \hspace{\tab} Man & 25 \\
    \hspace{\tab} Non-binary & 3 \\
    \hspace{\tab} Prefer not to say & 1 \\
    \midrule
    \textbf{Age} & \\
    \hline
    \hspace{\tab} 35-44 & 26 \\
    \hspace{\tab} 25-34 & 19 \\
    \hspace{\tab} 55-64 & 17 \\
    \hspace{\tab} 45-54 & 16 \\
    \hspace{\tab} 65 or older & 7 \\
    \hspace{\tab} 18-24 & 7 \\
    \midrule
    \textbf{Visual Acuity} & \\
    \hline
    \hspace{\tab} Totally blind (no light or shape perception) & 38 \\
    \hspace{\tab} Legally blind, with only light perception & 18 \\
    \hspace{\tab} Legally blind, partial sight & 8 \\
    \hspace{\tab} Prefer to self-describe & 7 \\
    \hspace{\tab} Legally blind, with both light and shape perception & 6  \\
    \hspace{\tab} Legally blind, peripheral vision loss & 6 \\
    \hspace{\tab} Legally blind, central vision loss & 4 \\
     \hspace{\tab} Legally blind, fluctuating vision & 3 \\
    \hspace{\tab} Legally blind, blurry vision & 1 \\
    \hspace{\tab} Legally blind, tunnel vision & 1 \\
    \hspace{\tab} Legally blind, with only shape perception & 1 \\
    \bottomrule
    \end{tabular}
    \caption{Demographic characteristics of survey respondents considered in final analysis (n=93).}
    \Description{Table showing demographic characteristics (gender, age, and visual acuity) of DMH tracking service users throughout the blind community (n=93). Women comprised the majority of participants' gender (64), followed by men (25), non-binary (3) and prefer not to say (1). The 35-44 age range was the most represented (26), followed by 25-34 (19), 55-64 (17), 45-54 (16), 65 or older (7), and 18-24 (7). Totally blind (no light or shape perception) was the most represented visual acuity (38), followed by legally blind - with only light perception (18), Legally blind -  partial sight (8), Prefer to self-describe (7), Legally blind - with both light and shape perception (6), Legally blind, peripheral vision loss (6), Legally blind - central vision loss (4), Legally blind - fluctuating vision (3), Legally blind - blurry vision (1), Legally blind - tunnel vision (1), and Legally blind - with only shape perception (1).}
    \label{tab:survey-demographics}
    \hfill
\end{table}

\begin{figure*}[h!b]
    \centering
        \begin{subfigure}[b]{0.8\textwidth}
        \centering
        \includegraphics[width=\textwidth]{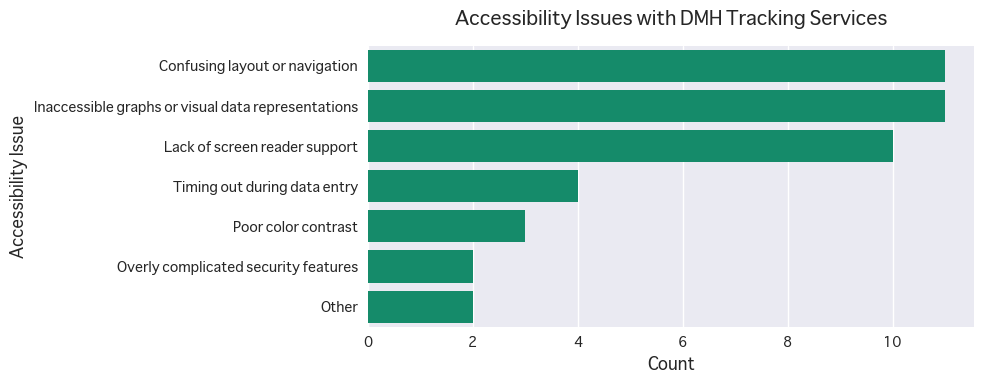}
        \caption{Reported access challenges when using DMH tracking services.}
        \Description{Horizontal bar chart titled ‘Accessibility Issues with DMH tracking services’, listing the following issues with their respective counts: ‘Confusing layout or navigation’ (11), ‘Inaccessible graphs or visual data representations’ (11), ‘Lack of screen reader support’ (9), ‘Timing out during data entry’ (6), ‘Poor color contrast’ (4), ‘Overly complicated security features’ (2), and ‘Other’ (2).}
        \label{subfig:access-issues}
    \end{subfigure}
    \hfill
    \begin{subfigure}[b]{0.8\textwidth}
        \centering
        \includegraphics[width=\textwidth]{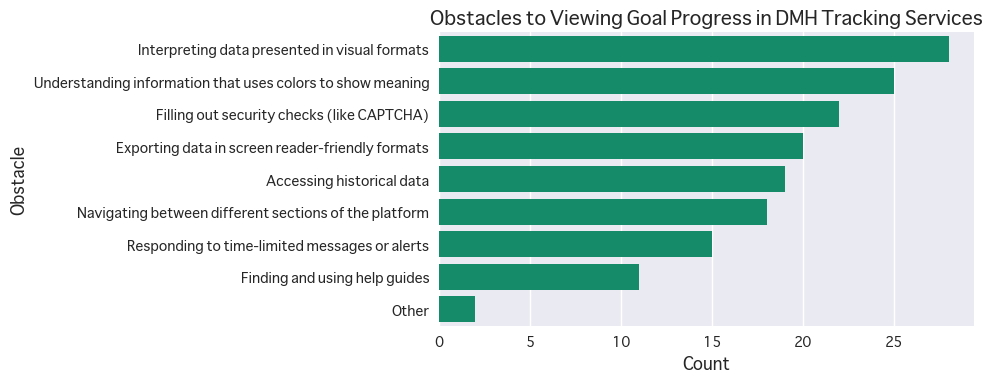}
        \caption{Reported obstacles to viewing goal progress with DMH tracking services.}
        \Description{Horizontal bar chart titled ‘Obstacles to Viewing Goal Progress in DMH Tracking Services’, listing the following obstacles with their respective counts: ‘Interpreting data presented in visual formats’ (25), ‘Understanding information that uses colors to show meaning’ (22), ‘Filling out security checks (like CAPTCHA)’ (18), ‘Exporting data in screen reader-friendly formats’ (16), ‘Accessing historical data’ (15), ‘Navigating between different sections of the platform’ (14), ‘Responding to time-limited messages or alerts’ (10), ‘Finding and using help guides’ (9), and ‘Other’ (2).}
        \label{subfig:obstacles}
    \end{subfigure}
    \hfill
    \begin{subfigure}[b]{0.8\textwidth}
        \centering
        \includegraphics[width=\textwidth]{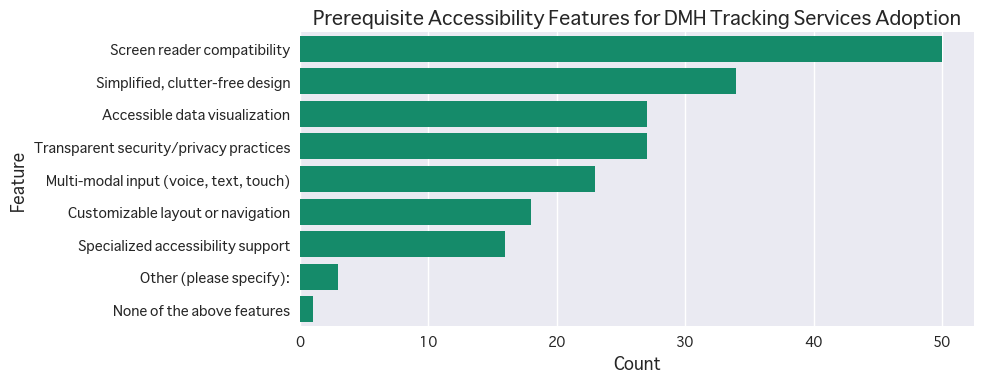}
        \caption{Reported prerequisite access features for DMH tracking service adoption.}
        \Description{Horizontal bar chart showing prerequisite accessibility features for DMH tracking services adoption. Y-axis lists features, X-axis shows count. Features in descending order: Screen reader compatibility (50), Simplified clutter-free design (40), Accessible data visualization (30), Transparent security/privacy practices (30), Multi-modal input (25), Customizable layout or navigation (20), Specialized accessibility support (15), Other (5), None of the above features (2).}
        \label{subfig:prereq-features}
    \end{subfigure}
    
    \caption{Participants' specific usability challenges and requirements for DMH tracking services.}
    \Description{Three horizontal bar charts showing participants' Reported obstacles to viewing goal progress(left), reported access challenges (center), and reported prerequisite accessibility features before DMH tracking service adoption (right).}
    \label{fig:dmh-specific-issues}
\end{figure*}

\clearpage

\section{Survey Instrument}
\label{sec:survey-instrument}

\begin{enumerate}
    \item When was the last time you used a digital mental health tracking service?
    \begin{enumerate}
        \item Today
        \item This past week 
        \item This past month 
        \item This past year
        \item I have never used DMH tracking services.
    \end{enumerate}

    \subsection*{For Current Users and Those Planning to Use Services}
    \item[T1/T2.1] Which of the following digital mental health tracking services have you used before/do you plan to use?
    \begin{enumerate}
        \item[\( \square \)] Cognitive and mental exercise services (e.g., Lumosity, Happify) 
        \item[\( \square \)] Habit and goal tracking services (e.g., Habitica, Strides) 
        \item[\( \square \)] Medication and symptom management (e.g., Medisafe, Bearable)  
        \item[\( \square \)] Mindfulness and meditation services (e.g., Headspace, Calm)
        \item[\( \square \)] Mood and emotion tracking services (e.g., Daylio, Moodfit)  
        \item[\( \square \)] Self-care and lifestyle management (e.g., Fabulous, Shine)  
        \item[\( \square \)] Sleep tracking services (e.g., Sleep Cycle, Fitbit) 
        \item[\( \square \)] Stress and anxiety management services (e.g., Sanvello, Pacifica, What's Up?) 
        \item[\( \square \)] Support community services (e.g., 7 Cups, NAMI)  
        \item[\( \square \)] Therapy and counseling services (e.g., MoodTools, Woebot)  
    \end{enumerate}

    \item[T1/T2.2] Of the digital mental health tracking services that you have used/that you plan to use, when you have used them, how often did you use them/how often will you use them (see Table~\ref{tab:service-matrix})~\footnote{1 = Never, 2 = Rarely, 3 = Once a month, 4 = Several times a month, 5 = Once a week, 6 = Several times a week, 7 = Daily.}?
\begin{table*}[t]
    \centering
    \caption{Matrix of service types from which participants would indicate their frequency of use.}
    \label{tab:service-matrix}
    \begin{tabular}{|l|c|c|c|c|c|c|c|}
    \hline
    \textbf{Service} & \textbf{1} & \textbf{2} & \textbf{3} & \textbf{4} & \textbf{5} & \textbf{6} & \textbf{7} \\
    \hline
    Cognitive and mental exercise services & \(\bigcirc\) & \(\bigcirc\) & \(\bigcirc\) & \(\bigcirc\) & \(\bigcirc\) & \(\bigcirc\) & \(\bigcirc\) \\
    Habit and goal tracking services & \(\bigcirc\) & \(\bigcirc\) & \(\bigcirc\) & \(\bigcirc\) & \(\bigcirc\) & \(\bigcirc\) & \(\bigcirc\) \\
    Medication and symptom management & \(\bigcirc\) & \(\bigcirc\) & \(\bigcirc\) & \(\bigcirc\) & \(\bigcirc\) & \(\bigcirc\) & \(\bigcirc\) \\
    Mindfulness and meditation services & \(\bigcirc\) & \(\bigcirc\) & \(\bigcirc\) & \(\bigcirc\) & \(\bigcirc\) & \(\bigcirc\) & \(\bigcirc\) \\
    Mood and emotion tracking services & \(\bigcirc\) & \(\bigcirc\) & \(\bigcirc\) & \(\bigcirc\) & \(\bigcirc\) & \(\bigcirc\) & \(\bigcirc\) \\
    Self-care and lifestyle management & \(\bigcirc\) & \(\bigcirc\) & \(\bigcirc\) & \(\bigcirc\) & \(\bigcirc\) & \(\bigcirc\) & \(\bigcirc\) \\
    Sleep tracking services & \(\bigcirc\) & \(\bigcirc\) & \(\bigcirc\) & \(\bigcirc\) & \(\bigcirc\) & \(\bigcirc\) & \(\bigcirc\) \\
    Stress and anxiety management services & \(\bigcirc\) & \(\bigcirc\) & \(\bigcirc\) & \(\bigcirc\) & \(\bigcirc\) & \(\bigcirc\) & \(\bigcirc\) \\
    Support community services & \(\bigcirc\) & \(\bigcirc\) & \(\bigcirc\) & \(\bigcirc\) & \(\bigcirc\) & \(\bigcirc\) & \(\bigcirc\) \\
    Therapy and counseling services & \(\bigcirc\) & \(\bigcirc\) & \(\bigcirc\) & \(\bigcirc\) & \(\bigcirc\) & \(\bigcirc\) & \(\bigcirc\) \\
    \hline
    \end{tabular}
    \Description{Matrix of service types from which participants would indicate their frequency of use.}
\end{table*}

    \item[T1/T2.3] Please list which specific digital mental health tracking services you have used/plan to use.

    \item[T1/T2.4] What were the high-level reasons you decided to start using digital mental health tracking services?
    \begin{enumerate}
        \item[\( \square \)] Finding support groups/communities
        \item[\( \square \)] Goal setting and/or tracking  
        \item[\( \square \)] Habit formation 
        \item[\( \square \)] Medication and symptom management 
        \item[\( \square \)] Mood tracking 
        \item[\( \square \)] Practicing mindfulness and meditation 
        \item[\( \square \)] Self-care and lifestyle management 
        \item[\( \square \)] Stress and anxiety management
        \item[\( \square \)] Sleep tracking  
        \item[\( \square \)] Therapy and counseling 
        \item[\( \square \)] Other (please specify): 
    \end{enumerate}

    \item[T1/T2.5] If comfortable sharing, please elaborate on these reasons to help us understand why you decided to start using these services. 

    \item[T1/T2.6] To what extent have digital mental health tracking services been helpful/do you anticipate them finding helpful in managing your mental health needs?
    \begin{enumerate}
        \item[\(\bigcirc\)] Very unhelpful 
        \item[\(\bigcirc\)] Somewhat unhelpful
        \item[\(\bigcirc\)] Neither helpful nor unhelpful 
        \item[\(\bigcirc\)] Somewhat helpful
        \item[\(\bigcirc\)] Very helpful  
    \end{enumerate}

    \item[T1/T2.7] What did you find/anticipate finding helpful or unhelpful about these services? 

    \item[T1/T2.8] How accessible are digital mental health tracking services for you when using tools like assistive technologies?
    \begin{enumerate}
        \item[\(\bigcirc\)] Very inaccessible 
        \item[\(\bigcirc\)] Somewhat inaccessible
        \item[\(\bigcirc\)] Neither accessible nor inaccessible 
        \item[\(\bigcirc\)] Somewhat accessible
        \item[\(\bigcirc\)] Very accessible  
    \end{enumerate}

    \item[T2.9] What alternative methods, if any, do you use to track your mental health?
    \begin{enumerate}
        \item[\( \square \)] In-person therapy or counseling services 
        \item[\( \square \)] Journaling
        \item[\( \square \)] In-person support groups
        \item[\( \square \)] Manual tracking systems
        \item[\( \square \)] Medication and treatment diaries 
        \item[\( \square \)] Physical activity 
        \item[\( \square \)] Self-assessment questionnaires
        \item[\( \square \)] Spiritual and/or religious practices
        \item[\( \square \)] Other (please specify):
        \item[\( \square \)] I do not currently use any strategies to manage my mental health.  
    \end{enumerate}

    \item[T1/T2.10] Which of the following accessibility issues have you encountered when using digital mental health tracking services?
    \begin{enumerate}
        \item[\( \square \)] Confusing layout or navigation 
        \item[\( \square \)] Inaccessible graphs or visual data representations 
        \item[\( \square \)] Lack of screen reader support   
        \item[\( \square \)] Overly complicated security features   
        \item[\( \square \)] Poor color contrast   
        \item[\( \square \)] Timing out during data entry   
        \item[\( \square \)] Other (please specify):   
    \end{enumerate}
    
    \item[T1/T2.11] How accessible are digital mental health tracking services for you \textbf{from a financial perspective}?
    \begin{enumerate}
        \item[\(\bigcirc\)] Very inaccessible 
        \item[\(\bigcirc\)] Somewhat inaccessible
        \item[\(\bigcirc\)] Neither accessible nor inaccessible 
        \item[\(\bigcirc\)] Somewhat accessible
        \item[\(\bigcirc\)] Very accessible  
    \end{enumerate}
    
    \item[T1/T2.12] Please describe your overall experiences with accessibility on digital mental health tracking platforms (technical, financial, etc.). 

    \item[T1/T2.13] From your experience, how effective have digital mental health tracking services been at meeting your mental health goals?
    \begin{enumerate}
        \item[\(\bigcirc\)] Very ineffective 
        \item[\(\bigcirc\)] Somewhat ineffective
        \item[\(\bigcirc\)] Neither effective nor ineffective 
        \item[\(\bigcirc\)] Somewhat effective
        \item[\(\bigcirc\)] Very effective 
    \end{enumerate}

    \item[T1/T2.14] Please describe what made these services ineffective or effective for you and why. 
    
    \item[T1/T2.15] How difficult is it for you to track your progress towards achieving mental health goals using digital mental health tracking platforms?
    \begin{enumerate}
        \item[\(\bigcirc\)] Very difficult 
        \item[\(\bigcirc\)] Somewhat difficult
        \item[\(\bigcirc\)] Neither easy nor difficult 
        \item[\(\bigcirc\)] Somewhat easy
        \item[\(\bigcirc\)] Very easy 
    \end{enumerate}
    
    \item[T1/T2.16] What assistive technologies, if any, do you typically use when interacting with digital mental health tracking services?
    \begin{enumerate}
        \item[\( \square \)] Accessibility and/or usability barriers
        \item[\( \square \)] Braille display 
        \item[\( \square \)] Customized app/browser extensions or add-ons
        \item[\( \square \)] Keyboard navigation  
        \item[\( \square \)] Magnification software 
        \item[\( \square \)] Refreshable Braille keyboard 
        \item[\( \square \)] Screen customization (e.g., high contrast, text size) 
        \item[\( \square \)] Screen reader  
        \item[\( \square \)] Voice commands 
        \item[\( \square \)] Other (please specify): 
        \item[\( \square \)] I do not use any assistive technologies when interacting with these services. 
    \end{enumerate}

    \item[T1/T2.17] What challenges, if any, do you encounter when trying to view your goal progress in a digital mental health tracking service?
    \begin{enumerate}
        \item[\( \square \)] Accessing historical data  
        \item[\( \square \)] Exporting data in screen reader-friendly formats 
        \item[\( \square \)] Filling out security checks (like CAPTCHA)  
        \item[\( \square \)] Finding and using help guides 
        \item[\( \square \)] Interpreting data presented in visual formats 
        \item[\( \square \)] Navigating between different sections of the platform 
        \item[\( \square \)] Responding to time-limited messages or alerts 
        \item[\( \square \)] Understanding information that uses colors to show meaning 
        \item[\( \square \)] Other (please specify):  
        \item[\( \square \)] I do not face any challenges while using digital mental health tracking services. 
    \end{enumerate}

    \item[T1/T2.18] What features or accommodations would need to be present for you to consider using a digital mental health tracking service?
    \begin{enumerate}
        \item[\( \square \)] Accessible data visualization 
        \item[\( \square \)] Customizable layout or navigation 
        \item[\( \square \)] Multi-modal input (voice, text, touch) 
        \item[\( \square \)] Screen reader compatibility 
        \item[\( \square \)] Simplified, clutter-free design  
        \item[\( \square \)] Specialized accessibility support 
        \item[\( \square \)] Transparent security/privacy practices
        \item[\( \square \)] Other (please specify): 
    \end{enumerate}
    
    \item[T1/T2.19] Could you please briefly elaborate on the challenges you mentioned or describe any other challenges you face with digital mental health tracking services? 

    \item[T1/T2.20] Could you please describe why you don't face any challenges with these services? 
    
    \item[T1/T2.21] Are there any features of the digital mental health tracking services that you have had experience with that have made you feel in control/might make you feel in control of your data on the service? If so, please briefly describe what these features are. If not, please write "N/A".
    
    \item[T1/T2.22] Do you have any privacy or security concerns about using digital mental health platforms? If so, please briefly describe them. If not, please briefly describe why that is the case.

    \subsection*{For Those Not Planning to Use Services}
    \item[T3.1] Please indicate why you do not plan on using digital mental health tracking services in the future or why you have stopped using them.
    \begin{enumerate}
        \item[\( \square \)] Accessibility and/or usability barriers
        \item[\( \square \)] Cost of use 
        \item[\( \square \)] Lack of awareness about these services  
        \item[\( \square \)] Preference for traditional, in-person methods
        \item[\( \square \)] Security/privacy concerns 
        \item[\( \square \)] Skepticism about effectiveness
        \item[\( \square \)] Time constraints (not enough time to learn a new technology) 
        \item[\( \square \)] Other (please specify): 
    \end{enumerate}

    \item[T3.2] If comfortable sharing, please describe your reasons below to help us better understand your reasons for not wanting to use digital mental health tracking services.

    \item[T3.3] What alternative methods, if any, do you use to track your mental health?
    \begin{enumerate}
        \item[\( \square \)] In-person therapy or counseling services 
        \item[\( \square \)] Journaling
        \item[\( \square \)] In-person support groups
        \item[\( \square \)] Manual tracking systems
        \item[\( \square \)] Medication and treatment diaries 
        \item[\( \square \)] Physical activity 
        \item[\( \square \)] Self-assessment questionnaires
        \item[\( \square \)] Spiritual and/or religious practices
        \item[\( \square \)] Other (please specify):
        \item[\( \square \)] I do not currently use any strategies to manage my mental health.  
    \end{enumerate}

        \item[T3.4] If comfortable sharing, please describe how you currently manage your mental health goals without using digital mental health tracking services.

    \item[T3.5] What strategies, if any, are you interested in trying to achieve your mental health goals?

    \item[T3.6] Have you had any negative experiences with other digital health services in general that have influenced your decision not to use digital mental health tracking services?
    \begin{enumerate}
        \item[\(\bigcirc\)] Yes
        \item[\(\bigcirc\)] No
        \item[\(\bigcirc\)] I have not used other digital health services.
    \end{enumerate}

    \item[T3.7] If comfortable sharing, could you briefly describe these experiences to help us better understand your experiences with digital mental health tracking services?

    \item[T3.8] What features or accommodations would need to be present for you to consider using a digital mental health tracking service?
    \begin{enumerate}
        \item[\( \square \)] Accessible data visualization 
        \item[\( \square \)] Customizable layout or navigation 
        \item[\( \square \)] Multi-modal input (voice, text, touch) 
        \item[\( \square \)] Screen reader compatibility 
        \item[\( \square \)] Simplified, clutter-free design  
        \item[\( \square \)] Specialized accessibility support 
        \item[\( \square \)] Transparent security/privacy practices
        \item[\( \square \)] Other (please specify): 
    \end{enumerate}

    \item[T3.9] Please explain why your selected features might influence your decision to use digital mental health tracking services.

    \item[T3.10] How important is data privacy and control to you when considering mental health management tools?
    \begin{enumerate}
        \item[\(\bigcirc\)] Extremely important  
        \item[\(\bigcirc\)] Very Important
        \item[\(\bigcirc\)] Moderately important  
        \item[\(\bigcirc\)] Slightly important
        \item[\(\bigcirc\)] Not at all important  
    \end{enumerate}

    \item[T3.11] Please describe why data privacy and control with mental health management tools is (or isn't) important to you.

    \subsection*{Demographics}
    \item[D.1] What is your age in years?
    \begin{enumerate}
        \item[\(\bigcirc\)] 18-24
        \item[\(\bigcirc\)] 25-34 
        \item[\(\bigcirc\)] 35-44 
        \item[\(\bigcirc\)] 45-54 
        \item[\(\bigcirc\)] 55-64 
        \item[\(\bigcirc\)] 65 or older
    \end{enumerate}

    \item[D.2] What is your gender?
    \begin{enumerate}
        \item[\(\bigcirc\)] Woman
        \item[\(\bigcirc\)] Man
        \item[\(\bigcirc\)] Non-binary
        \item[\(\bigcirc\)] Prefer to self-describe
        \item[\(\bigcirc\)] Prefer not to say
    \end{enumerate}

    \item[D.3] What is your level of visual acuity?
    \begin{enumerate}
        \item[\(\bigcirc\)] Totally blind (no light or shape perception)  
        \item[\(\bigcirc\)] Legally blind, with both light and shape perception  
        \item[\(\bigcirc\)] Legally blind, with only light perception  
        \item[\(\bigcirc\)] Legally blind, with only shape perception  
        \item[\(\bigcirc\)] Legally blind, central vision loss  
        \item[\(\bigcirc\)] Legally blind, peripheral vision loss  
        \item[\(\bigcirc\)] Legally blind, tunnel vision  
        \item[\(\bigcirc\)] Legally blind, blurry vision  
        \item[\(\bigcirc\)] Legally blind, fluctuating vision  
        \item[\(\bigcirc\)] Legally blind, partial sight 
        \item[\(\bigcirc\)] Prefer to self-describe 
    \end{enumerate}

    \subsection*{Follow-up Contact}
    \item[F.1] Are you interested in being contacted about a paid follow-up interview with a graduate student researcher from the research team?
    \begin{enumerate}
        \item[\(\bigcirc\)] No
        \item[\(\bigcirc\)] Yes
    \end{enumerate}

    \item[F.2] Please enter your preferred email address so that we may contact you for a follow-up interview.

\end{enumerate}

\end{document}